\documentclass[12pt]{article}
\usepackage{latexsym}
\usepackage{graphics}

\begin{document}

%\begin{flushright}
%Sofia University\\
%\end{flushright}
%%%%%%%%%%%%%%%%%%%%%%%%%%%%%%%%%%%%%%%%%%%%%%%%%%%%%%%%%%%%%%%%%%%

\title{Scalar-tensor  black holes  coupled to Born-Infeld nonlinear
electrodynamics}
\author{Ivan Zh. Stefanov\thanks{E-mail: zhivkov@phys.uni-sofia.bg}, \,\,
     Stoytcho S. Yazadjiev \thanks{E-mail: yazad@phys.uni-sofia.bg;
        \emph{Address since 1st April 2007: Institut f\"{u}r Theoretische Physik,
         Universitat Gottingen,
        Friedrich-Hund-Platz 1, D-37077 G\"{o}ttingen, Germany} }\\
{\footnotesize  Dept. of Theoretical Physics,
                Faculty of Physics}\\ {\footnotesize St.Kliment Ohridski University of Sofia}\\
{\footnotesize  5, James Bourchier Blvd., 1164 Sofia, Bulgaria }\\\\[-3.mm]
 Michail D.~Todorov\thanks{E-mail: mtod@tu-sofia.bg}
\\ [-1.mm]{\footnotesize
{Faculty of Applied Mathematics and Computer Science}}\\
[-1.mm] {\footnotesize {Technical University of Sofia}}\\
[-1.mm] {\footnotesize 8, Kliment Ohridski Blvd., 1000 Sofia, Bulgaria}}

\date{}

\maketitle

\begin{abstract}
The non-existence of asymptotically flat, neutral black holes and asymptotically flat, charged black holes in the
Maxwell electrodynamics,
with non-trivial scalar field has been proved for a large class of scalar-tensor theories. The no-scalar-hair theorems,
however, do not apply in the case of non-linear electrodynamics. In the present work numerical solutions describing
charged black holes coupled to Born-Infeld type non-linear electrodynamics in scalar-tensor theories of gravity
with massless scalar field are found. The causal structure and properties of the solutions are studied, and a comparison
between these solutions and the corresponding solutions in the General Relativity is made. The presence of the scalar field
leads to a much more simple causal structure. The present class of black holes has a single, non-degenerate horizon, i.e.,
its causal structure resembles that of the Schwarzschild black hole.
\end{abstract}

%%%%%%%%%%%%%%%%%%%%%%%%%%%%%%%%%%%%%%%%%%%%%%%%%%%%%%%%%%%%%%%%%%%

%\draft
\sloppy

\section{Introduction}

Scalar-tensor theories of gravity are the most natural generalization of General  Relativity (GR)
and arise naturally in string theory and higher dimensional gravity theories \cite{DP}. Different modifications of
scalar-tensor theories are attracting much interest also in cosmology and astrophysics. The most natural question is
whether objects found and studied in the frame of GR %General relativity
would have different properties in the
frame of the scalar-tensor theories. Such objects of interests are the black holes.
A number of no-hair theorems concerning their global properties have been proved. The no-hair conjecture in GR %General relativity
says
that in the exterior of a black hole
the only information available regarding the black hole may be that of its mass, charge, and angular momentum.
For neutral static, spherically symmetric black holes the only available information is about its mass, and the
exterior of the black hole is reduced to the Schwarzschild solution.
The natural question which arises then is whether the scalar field would lead to the existence of other preserved
quantities which would allow a distant observer to distinguish between the Schwarzschild black hole
and a black hole with a scalar dressing.
Saa \cite{Saa} was able to prove a no-scalar-hair theorem  which rules out the existence of static, spherically symmetric,
asymptotically flat, neutral
black holes with regular, non-trivial scalar field for a large class of scalar tensor theories in which the scalar field
is non-minimally coupled to gravity. He applied an explicit, covariant method to generate the exterior solutions for
these theories  through conformal transformations from the minimally coupled case. The
scalar field in these theories becomes a constant and thus trivial if one demands that the essential
singularity at the center of symmetry is hidden by an event horizon. A similar theorem treating also the
case of charged scalar field with self-interaction was proved by Bekenstein \cite{Be}.
Saa's theorem was generalized for the case of charged black holes in linear electrodynamics by Banerjee and Sen
\cite{BSen}.

In the case of non-linear electrodynamics, however, the energy-momentum tensor of the electromagnetic field
has a non-zero trace a sequence of which is that the electromagnetic field is non-trivially coupled to the scalar field. Hence,
we can expect that the no-scalar-hair theorems might not hold in that case. In the present work we prove that
our assumption is correct
and find numerical solutions describing black holes with a non-trivial scalar field in the non-linear electrodynamics.

%In this paper we consider a particular example of nonlinear electrodynamics, the Born-Infeld electrodynamics.
The nonlinear electrodynamics was first introduced by Born and Infeld in 1934 to obtain finite energy density
model for the electron \cite{BI}.
They proposed the following Lagrangian
\begin{equation}
L_{BI} = 2b \left[ 1- \sqrt{1+ \frac{1}{4b}F_{\mu\nu}F^{\mu\nu}-\frac{1}{64b^2}(F_{\mu\nu}\star F^{\mu\nu})^2 } \right],
\end{equation}
where the star ``$\star$'' stands for the Hodge operator.
In recent years nonlinear electrodynamics models are attracting much interest, too.
The reason is that the nonlinear electrodynamics arises naturally in open strings and $D$-branes \cite{L}.
Nonlinear electrodynamics models coupled to gravity have been discussed  in different aspects
(see, for example, \cite{Demianski}--\cite{Iran}
and references therein).

%Here we obtain and discuss black hole solutions coupled to the nonlinear Born-Infeld electrodynamics within
%scalar-tensor theories. Black hole solutions which are neutral or coupled to the linear electrodynamics
%in the presence of a dilaton field were discussed by many authors \cite{MW1}-\cite{Y}.

%Asymptotically flat black
%holes in Born-Infeld theory coupled to Einstein gravity were studied in \cite{}.

%The effective Lagrangian for electrodynamics due to one-loop quantum corrections was calculated by Heisenberg and Euler
%\cite{EH}:
%
%\begin{equation}
%L_{EH} = -{1\over 4}F_{\mu\nu}F^{\mu\nu} + \frac{1}{4} \,b^2 \left(F_{\mu\nu}F^{\mu\nu}\right)^2 + \gamma \left[ F_{\mu\nu}(\star F)^{\mu\nu} \right]^2
%\end{equation}
%
%where $b^2= {8he^4/ (2880 \pi^2 m^4)}$, $\gamma = {7he^4/(5760\pi^2m^4)}$ and $h$, $e$ and $m$ are the Planck constant,
%electron charge, and electron mass, respectively. The star sign denotes the Hodge dual. From experimental aspect,
%the Euler-Heisenberg theory is more accurate classical approximation of QED than the Maxwell theory when the field has
%high intensity\cite{SB}. Regarding the electric-magnetic duality, the Euler-Heisenberg action breaks it as it was
%pointed in \cite{GR1}.

\section{Formulation of the problem}

The general form of the extended gravitational action in
scalar-tensor theories is

\begin{eqnarray} \label{JFA}
S = {1\over 16\pi G_{*}} \int d^4x \sqrt{-{\tilde
g}}\left({F(\Phi)\tilde {\cal R}} - Z(\Phi){\tilde
g}^{\mu\nu}\partial_{\mu}\Phi
\partial_{\nu}\Phi  \right. \nonumber  \\  -2 U(\Phi) \Bigr) +
S_{m}\left[\Psi_{m};{\tilde g}_{\mu\nu}\right] .
\end{eqnarray}

Here, $G_{*}$ is the bare gravitational constant, ${\tilde R}$ is
the Ricci scalar curvature with respect to the space-time metric
${\tilde g}_{\mu\nu}$. The dynamics of the scalar field $\Phi$
depends on the functions $F(\Phi)$, $Z(\Phi)$ and $U(\Phi)$. In
order for the gravitons to carry positive energy the function
$F(\Phi)$ must be positive. The nonnegativity of the energy of
the dilaton field requires that $2F(\Phi)Z(\Phi) +
3[dF(\Phi)/d\Phi]^2 \ge 0$. The action of matter depends on the
material fields $\Psi_{m}$ and the space-time metric ${\tilde
g}_{\mu\nu}$. It does not involve the scalar field $\Phi$ in
order for the weak equivalence principle to be satisfied.

It is much more convenient from a mathematical point of
view to analyze the scalar-tensor theories with respect to the
conformally  related Einstein frame  given by the metric:

\begin{equation}
g_{\mu\nu} = F(\Phi){\tilde g}_{\mu\nu}. \label {conf}
\end{equation}

Further, let us introduce the scalar field $\varphi$ (the so-called dilaton) via the equation

\begin{equation}\label {CONF2}
\left(d\varphi \over d\Phi \right)^2 = {3\over
4}\left\{{d\ln[F(\Phi)]\over d\Phi } \right\}^2 + {Z(\Phi)\over 2
F(\Phi)}
\end{equation}

 and define

\begin{equation}\label {CONF3}
{\cal A}(\varphi) = F^{-1/2}(\Phi) \,\,\, ,\nonumber \\
2V(\varphi) = U(\Phi)F^{-2}(\Phi) .
\end{equation}

In the Einstein frame action (\ref{JFA}) takes the form

\begin{eqnarray}\label{EFA}
S= {1\over 16\pi G_{*}}\int d^4x \sqrt{-g} \left[{\cal R} -
2g^{\mu\nu}\partial_{\mu}\varphi \partial_{\nu}\varphi -
4V(\varphi)\right] \nonumber \\ + S_{m}[\Psi_{m}; {\cal
A}^{2}(\varphi)g_{\mu\nu}]
\end{eqnarray}

where $R$ is the Ricci scalar curvature with respect to the
Einstein metric $g_{\mu\nu}$.

We take the following Jordan frame nonlinear electrodynamics action

\begin{equation}
S_{m} = {1\over 4\pi G_{*}}\int d^4x \sqrt{{-\tilde g}} L(X, Y)
\end{equation}

where

\begin{equation}
X = {1\over 4} F_{\mu\nu}{\tilde g}^{\mu\alpha} {\tilde g}^{\nu\beta} F_{\alpha\beta},  \,\,\,
Y = {1\over 4}  F_{\mu\nu}\left({\tilde \star} F\right)^{\mu\nu}
\end{equation}

and ``${\tilde \star}$" is the Hodge dual with respect to the Jordan frame metric ${\tilde g}_{\mu\nu}$.

In the Einstein frame we have

\begin{equation}\label{EFNEDA}
S_{m} = {1\over 4\pi G_{*}}\int d^4x \sqrt{-g} {\cal A}^4(\varphi) L(X, Y)
\end{equation}

where

\begin{equation}
X = {{\cal A}^{-4}(\varphi)\over 4} F_{\mu\nu}{g}^{\mu\alpha} {g}^{\nu\beta} F_{\alpha\beta}, \label{X}  \,\,\,
Y = {{\cal A}^{-4}(\varphi)\over 4}  F_{\mu\nu}\left({ \star} F\right)^{\mu\nu}
\end{equation}

and ``$\star$''  stands for the Hodge dual with respect to the Einstein frame metric $g_{\mu\nu}$.

The action (\ref{EFA}) with (\ref{EFNEDA}) yields the following field equations

\begin{eqnarray}
&&{\cal R}_{\mu\nu} = 2\partial_{\mu}\varphi \partial_{\nu}\varphi +  2V(\varphi)g_{\mu\nu} -
 2\partial_{X} L(X, Y) \left(F_{\mu\beta}F_{\nu}^{\beta} -
{1\over 2}g_{\mu\nu}F_{\alpha\beta}F^{\alpha\beta} \right)  \nonumber \\
&&-2{\cal A}^{4}(\varphi)\left[L(X,Y) -  Y\partial_{Y}L(X, Y) \right] g_{\mu\nu}, \nonumber  \\
&&\nabla_{\mu} \left[\partial_{X}L(X, Y) F^{\mu\nu} + \partial_{Y}L(X, Y) (\star F)^{\mu\nu} \right] = 0 \label{F},\\
&&\nabla_{\mu}\nabla^{\mu} \varphi = {dV(\varphi)\over d\varphi } -
4\alpha(\varphi){\cal A}^{4}(\varphi) \left[L(X,Y) -  X\partial_{X}L(X,Y) -  Y\partial_{Y}L(X, Y) \right], \nonumber
\end{eqnarray}
where $\alpha(\varphi) = {d \, \ln {\cal A}(\varphi)\over d\varphi}$.

In what follows we consider the truncated\footnote{Here we consider the pure magnetic case for which $Y=0$.  }
Born-Infeld  electrodynamics described by the Lagrangian

\begin{equation}
L_{BI}(X) = 2b \left( 1- \sqrt{1+ \frac{X}{b}} \right)\label{LBI}.
\end{equation}

Here $V(\varphi)$ will be equal to zero.
%\section{Magnetically charged black holes}

\section{Basic equations and qualitative investigation}

The metric of a static, spherically symmetric spacetime can be written in the form

\begin{equation}
ds^2 = g_{\mu\nu}dx^{\mu}dx^{\nu} = - f(r)e^{-2\delta(r)}dt^2 + {dr^2\over f(r) } +
r^2\left(d\theta^2 + \sin^2\theta d\phi^2 \right).
\end{equation}
In the considered Born-Infeld type non-linear electrodynamics an electric-magnetic duality exists which means that
the solutions in the magnetically charged case and the electrically charged case coincide. We will study the magnetically
charged black holes for which the electromagnetic field is given by

\begin{equation}
F = P \sin\theta d\theta \wedge d\phi
\end{equation}
and the magnetic charge is denoted by $P$.

The field equations reduce to the following coupled system of
ordinary differential equations:
\begin{eqnarray}
&&\frac{d\delta}{dr}=-r\left(\frac{d\varphi}{dr} \right)^2\label{EQDelta},\\
&&\frac{d m}{dr}=r^2\left[\frac{1}{2}f\left(\frac{d\varphi}{dr} \right)^2 - {\cal A (\varphi)}^{4}L(X)  \right] \label{EQm},\\
&&\frac{d }{dr}\left( r^{2}f\frac{d\varphi }{dr} \right)=
r^{2}\left\{-4\alpha(\varphi){\cal A}^{4}(\varphi) \left[L(X) -  X\partial_{X}L(X)\right] -
r f\left(\frac{d\varphi}{dr} \right)^3    \right\} \label{EQPhi}  ,
\end{eqnarray}
where $ X $ reduces to:
\begin{equation}
X = {{\cal A}^{-4}(\varphi)\over 2} \frac{P^2}{r^4}.
\end{equation}

In this paper, we will be searching for black hole solutions. We define black hole solutions as such solutions that have
an event horizon on which the dilaton field $\varphi$ is regular.We will also require that $0<{\cal A}(\varphi)<\infty$
for $r_{H}\leq r \leq \infty$, where $r_{H}$ is the radius of the horizon.
The latter condition ensures the regularity of the transition between the Einstein and the Jordan  conformal frames.
The diversity and the properties of the
solutions depend strongly on the choice of the functions ${\cal A}(\varphi)$ (respectively on $\alpha(\varphi)$).
In the present work we will consider only theories for which $\alpha(\varphi)$ has a fixed positive sign for all values
of $\varphi$. The manner of investigation of solutions within theories with negative $\alpha(\varphi)$ is similar.
Theories in which
the coupling function changes its sign are much more complicated (also from numerical point of view)
since in them some interesting effects like
bifurcation of solutions can appear, especially when $\alpha(\varphi)\sim\varphi$. Such solutions
are currently being studied by the authors and the results will be given elsewhere.

Some general properties of the solution can be derived through an analytical investigation of the equations.
We will consider nonlinear electrodynamics for which the following relation holds
\begin{equation}
 X\partial_{X}L(X)- L(X)>0\label{EDHAM}.
\end{equation}
The Born-Infeld Lagrangian (\ref{LBI}), belongs to the same class of nonlinear electrodynamics.

Using the following equation
\begin{equation}
\frac{d }{dr}\left( e^{-\delta}r^{2}f\frac{d\varphi }{dr} \right)=4 r^2 e^{-\delta} \alpha(\varphi)
{\cal A}^{4}(\varphi)\left[X\partial_{X}L(X)- L(X)\right]>0,\label{phianl} \\
\end{equation}
which is another form of equation (\ref{EQPhi}), we find that $e^{-\delta}r^{2}f\frac{d\varphi }{dr} $
increases monotonously and has one zero at most, if any.
In case a black hole exists, $e^{-\delta}r^{2}f\frac{d\varphi }{dr}$ vanishes on the horizon. On the other hand,
as we already noted, this expression is a monotonously increasing function of $r$.
Therefore $e^{-\delta}r^{2}f\frac{d\varphi }{dr}<0$ and
$e^{-\delta}r^{2}f\frac{d\varphi }{dr}>0$, inside and outside the horizon, respectively. In order for this to be true,
$\frac{d\varphi }{dr}$ must be positive, which means that $\varphi$ increases monotonously.

The non-existence of inner horizons can be proved in another way. Let us admit that more than one horizon exists.
Then, integrating equation (\ref{phianl})
in the interval $r\in[r_{-},r_{+}]$ where we denote he first inner and the outer horizons with $r_{-}$
and $r_{+}$, respectively, i.e.,
\begin{eqnarray}
&&\left. \left( e^{-\delta}r^{2}f\frac{d\varphi }{dr} \right) \right|_{r_{+}}-
\left.\left( e^{-\delta}r^{2}f\frac{d\varphi }{dr} \right) \right|_{r_{-}}\nonumber \\
&&\hspace{1cm}=4 \int\limits_{r_{-}}^{r_{+}}r^2 e^{-\delta} \alpha(\varphi)
{\cal A}^{4}(\varphi)\left[X\partial_{X}L(X)- L(X)\right] dr>0, \nonumber \\
\end{eqnarray}
and having in mind that $f(r_{-})=0=f(r_{+})$, we reach a contradiction,
which means that our admission is incorrect.
So if a black hole exists it will have a single horizon, i.e., its causal structure will be Schwarzschild-like.
In both conformal frames, inside the event horizon a space-like singularity is hidden.

The qualitative behavior of $\delta(r)$ can easily be seen from equation (\ref{EQDelta}).
It decreases monotonously with $r$.

\subsection{Numerical results}
The nonlinear system (\ref{EQDelta})-(\ref{EQPhi}) is
inextricably coupled and the event horizon $r_H$ is {\it a priori} unknown boundary. In order to be solved, it is recast as a equivalent first order system of ordinary differential equations.  Following the physical assumptions of the matter under consideration the asymptotic boundary conditions are set, i.e.,
$$\lim_{r \to \infty}m(r) =M \quad (M \>{\rm is\> the\> mass\> of\> the\> black\> hole\> in\> the\> Einstein\> frame}),$$
$$ \lim_{r \to \infty}\delta(r)=\lim_{r \to \infty}\varphi(r)=0.$$
At the horizon both the relationship $$f(r_H)=0$$
and
the regularization condition
$$\left.\left(\frac{df}{dr}\!\cdot\! \frac{d \varphi}{d r}\right)\right|_{r=r_H} =\left.
\left\{ 4 \alpha(\varphi) {\cal A}^4(\varphi) [X \partial_X L(X)-L(X) ]\right\}\right|_{r=r_H}$$ concerning the spectral quantity $r_H$ must be held.
For the treating the above posed boundary-value problem (BVP) the Continuous Analog of
Newton Method (see, for example \cite{gavurin},\cite{jidkov},\cite{YFBT}) is used. After an appropriate linearization the original BVP
is rendered to solving a vector two-point BVP. On a discrete level sparse (almost diagonal) linear algebraic
systems with regard to increments of sought functions $\delta(r)$, $m(r)$, and $\varphi(r)$ have to be inverted.

%For regions where the ratio $(P/M)^2$ is considerably below the critical value $ 24/25 $ the behavior of the solutions
We studied the parametric space for fixed value of the coupling parameter $\alpha=0.01$
(this value is close to the one established on the bases of experimental data)
and for several values of the magnetic charge. Constant coupling parameter corresponds to the Brans-Dicke theory.
The behavior of the solutions for any
$\alpha(\varphi)>0$ for which $\alpha(0)=0.01$ are qualitatively the same and quantitatively very close to
the case we studied as the numerical investigations confirm.

Figures (1)-(4) show the value of the scalar field on the event horizon, the radius of the
black hole in the Einstein and in the Jordan conformal frames, and the temperature of the horizon as a function of the
mass of the black hole, respectively.
For higher masses the behavior of the solutions
resembles that of the GR %General relativistic
case. When we decrease the mass for a fixed value of $P$, however,
the solutions start to deviate from
the GR %General relativistic
case considerably. In GR \cite{GR1} %General relativity
the properties of the solution depend on the value of
following quantity
\begin{equation}
 M'=M-\cal{E},
\end{equation}
where $M$ is the mass of the black hole and in our notations
\begin{equation}
{\cal{E}}=\frac{P^{ 3/2 }}{\sqrt{\displaystyle\frac {4 \pi}{ \sqrt{2b}}}}
\frac{\pi^{3/2}}{3\Gamma(\frac{3}{4})^{2}}.
\end{equation}
For $M'\geq0$ a single non-degenerate horizon exists. The case $M'<0$ resembles the Reissner-Nordstr\"{o}m
solution in which extremal black holes exist.
In the case under consideration, however,
the absolute value of the scalar field rises considerably and prevents the emergence
of a degenerate horizon. This can be clearly seen from the behavior of the graph representing the case $P=3.5$ in
Fig.(\ref{TMagn}). The temperature of the
horizon approaches zero but then suddenly rises. For these values of the mass $M$ and the magnetic charge $P$
in GR %General relativity
an extremal
solution is reached.
Far from the extremal solution the dilaton
becomes almost zero and remains constant with the increase of mass $M$.

It can also be noted in Fig.(\ref{RhM}) that the radius of the horizon turns to zero for a finite value of the mass $M$ of the black hole.
The mass-radius relation in the Jordan frame is shown in Fig.(\ref{RhM_J}). For the values of the parameters we work with
the difference between Figs. (\ref{RhM}) and (\ref{RhM_J}) is insignificant.
%The temperature, unlike the extremal solution, is non-zero and rises as the mass
%$M$ decreases.

\section{Thermodynamics}
For the solution we study, in the Einstein frame, the First Law (FL) of thermodynamics holds.
The formal derivation of this law can be seen in the work of Rasheed \cite{Rasheed}.
The presence of the scalar field leads to the existence of a new charge namely the dilaton charge which we define in the
following way
\begin{equation}
{\cal{D}} = - \left. r^2 \frac{d\varphi}{dr}\right|_{r\rightarrow \infty}. \label{dilatoncharge}
\end{equation}
This charge, however,
is not independent and can be determined unambiguously once the mass and the magnetic charge of the black hole and the
asymptotic value of the scalar field at infinity are known.
The relation between the charges can be seen if we integrate
equation (\ref{phianl}) from the radius of the horizon to infinity. We obtain
\begin{equation}
{\cal{D}}=4 \int\limits_{r_{H}}^{\infty} r^2 \alpha(\varphi)
{\cal A}^{4}(\varphi)\left[X\partial_{X}L(X)- L(X)\right]dr.
\end{equation}
Since in the asymptotically flat case the scalar field is
fixed and
vanishing at infinity the term coming from the scalar field also vanishes and
 the FL of thermodynamics is the same as in the GR
\begin{equation}
\delta M=T \delta S + \Psi_{H}\delta P,\label{FirstLaw}
\end{equation}
where $T$, $S$ and $P$ are the temperature, the entropy, and the magnetic charge of the black hole, respectively. The
quantity $\Psi$ conjugate to the magnetic charge  is the potential of the magnetic field which is given by the
following definition
\begin{equation}
H_{\mu}=\partial_{\mu}\Psi.\label{defH}
\end{equation}
On the other hand the magnetic field is defined as
\begin{equation}
H_{\mu}=-\star G_{\mu\nu}\xi^{\nu},
\end{equation}
where
\begin{equation}
G_{\mu\nu}=-\frac{1}{2} \frac{\partial L}{\partial F_{\mu\nu} },
\end{equation}
$\xi=\frac{\partial}{\partial t}$ is the Killing vector generating time translations
and ``$\star$'' is the Hodge star operator.

Originally we formulated the theory in Jordan frame but studied the solutions in the conformally related Einstein frame
not only for mathematical convenience but also because the FL of black hole thermodynamics is naturally
connected with the Einstein frame as this can be seen below.
%One would ask whether the thermodynamic properties of the solutions are the same in both conformal frames.

The temperature of the event horizon is invariant under
conformal transformations of the metric that are unity at infinity \cite{Jacobson}.
The properly defined entropy is also the same in both conformal frames.
It has been proved that in the Jordan frame the entropy of the black hole is not simply one forth of
the horizon area \cite{MVisser,FordRoman} as in the Einstein frame and needs to be generalized. The entropy in the
Jordan frame is defined
as:
\begin{equation}
S_{J}=\frac{1}{4G_{*}}\int d^2x \sqrt{-^{(2)}{\tilde g}}F(\Phi).
\end{equation}
Using relation (\ref{conf}) we find that
\begin{equation}
S_{J}=\frac{1}{4G_{*}}\int d^2x \sqrt{-^{(2)} g}=S_{E}=S.
\end{equation}
In the last two equations $^{(2)}{\tilde g}$ and $^{(2)} g$ are the determinants of the induced metrics on the horizon
in the Jordan
and in the Einstein frame, respectively.

%The properly defined entropy
%in the Jordan frame is equal to the entropy in the Einstein frame, which means that it is
%enough to study the thermodynamics of the solutions only in one of the conformal frames.
%, and $S_{J}$ and $S_{E}$ are the entropies in the Jordan frame and in the Einstein frame respectively.
%
The term in the FL (\ref{FirstLaw}) connected with the magnetic charge is also preserved under the conformal
transformations.

In order for the FL of thermodynamics to be satisfied in the Jordan frame the mass should be
properly chosen since
the Arnowitt-Deser-Misner (ADM) masses in both frames are not equivalent. It can be easily shown that the
ADM mass in the Jordan frame $M_{J}$ is related to both the
ADM mass in the Einstein frame $M$ and the dilaton charge ${\cal D}$ in the following way

\begin{equation}
M_{J}=M+\alpha\cal{D}.
\end{equation}
For the Jordan frame, the proper mass in the FL of thermodynamics is the ADM mass in the Einstein
frame $M$. Similarly, for boson and fermion stars the proper measure for the energy of the system is again the
ADM mass in the Einstein frame $M$. For more details on the subject
we would refer the reader to the works \cite{Lee,Shapiro,Whinnett,Yazadji}.

\section{Conclusion}

In the present work numerical solutions describing charged black holes coupled to non-linear electrodynamics in
the scalar-tensor theories with massless scalar field were found. Since an electric-magnetic duality is present,
in this work only purely magnetically case was studied. For the Lagrangian of the non-linear electrodynamics the
truncated Born-Infeld
Lagrangian was chosen and scalar-tensor theories with positive coupling parameter were considered. As a result of the
numerical and analytical investigation, some general properties of the solutions were found. Due to the presence of the
massless scalar field the found solutions have a single,
non-degenerate event horizon, i.e., their causal
structure resembles that of the Schwarzschild black hole and is simpler than the corresponding solution in GR.
In both conformal frames, inside the event horizon a space-like singularity is hidden.
Black hole thermodynamics was also discussed.

\section*{Acknowledgments} This work was partially supported by
the Bulgarian National Science Fund under Grant MUF 04/05 (MU 408) and the Sofia University Research Fund N60.
The final stage of this work was completed after S. Y. has started his visit in 
Institut f\"{u}r Theoretische Physik, Universitat G\"{o}ttingen as an Alexander von Humboldt research fellow so 
this author would like to thank Alexander von Humboldt Foundation for the financial support and the 
Institut f\"{u}r Theoretische Physik for the hospitality.

\begin{figure}[htbp]%
\vbox{ \hfil \scalebox{1.0}{ {\includegraphics{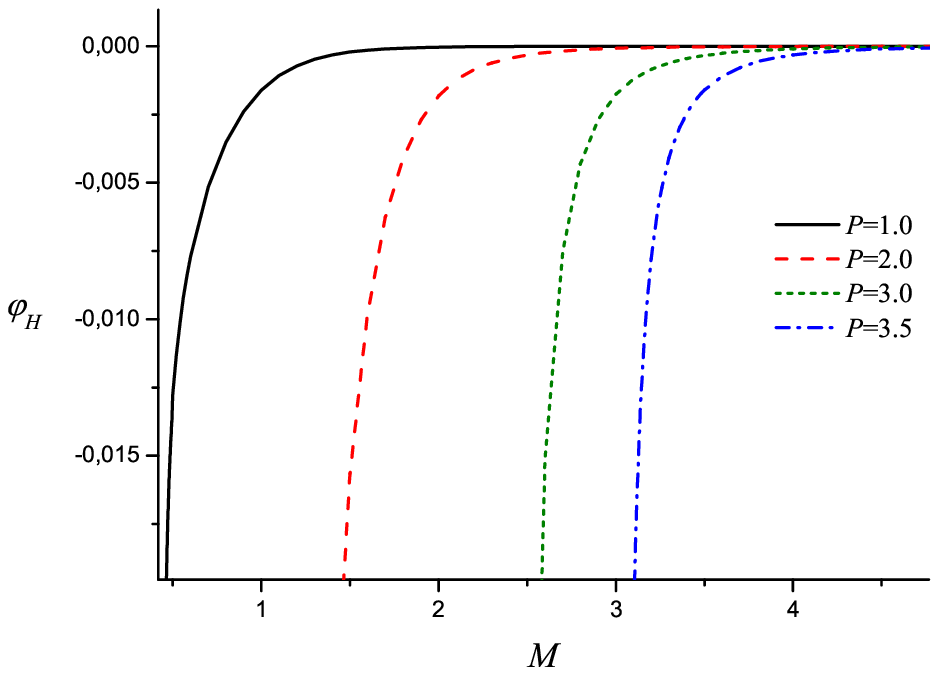}} }\hfil}%
\bigskip%
\caption{%
%------------------------------
The value of dilaton field $\varphi$ on the horizon as a function of the mass $M$, for
$P=1.0, 2.0, 3.0, 3.5 $ . For low values of $M$ the absolute value of the
dilaton increases considerably and prevents the formation of a degenerate horizon.} \label{PhiM}%
\end{figure}%

 \begin{figure}[htbp]%
\vbox{ \hfil \scalebox{1.0}{ {\includegraphics{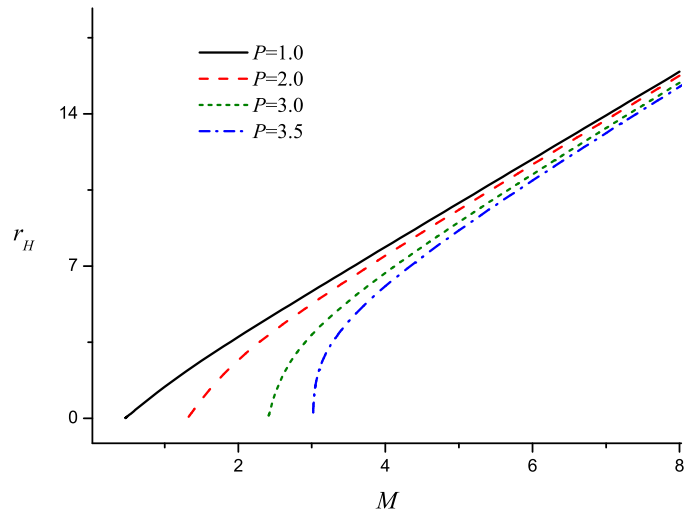}} }\hfil}%
\bigskip%
\caption{%
%------------------------------
The $ M$--$r_{H} $ relation for the same values of the parameters
as in  Fig.(\ref{PhiM}).
%------------------------------
} \label{RhM}%
\end{figure}%

 \begin{figure}[htbp]%
\vbox{ \hfil \scalebox{1.0}{ {\includegraphics{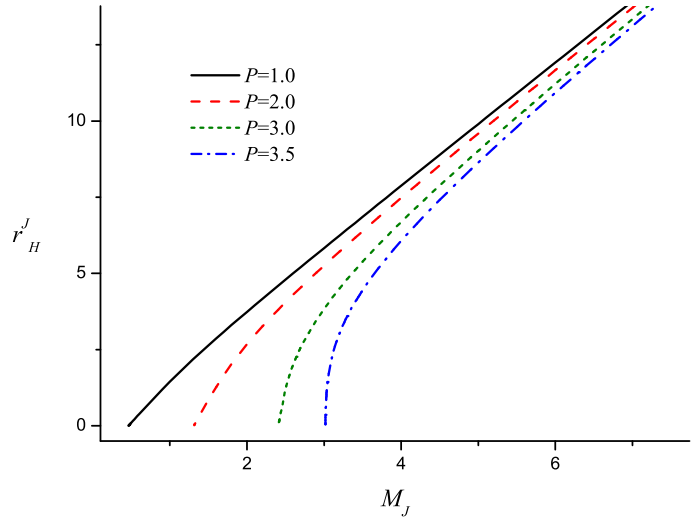}} }\hfil}%
\bigskip%
\caption{%
%------------------------------
The $ M_{J}-r^{J}_{H} $ relation, where $M_{J}$ and $r^{J}_{H}$ are respectively the mass and the radius of the black hole
in the Jordan frame, for the same values of the parameters
as in Fig.(\ref{PhiM}). For these values of the parameters (and $\alpha =0.01$)
the difference from the corresponding Fig.(\ref{RhM}) in the Einstein frame is insignificant.
%------------------------------
} \label{RhM_J}%
\end{figure}%

 \begin{figure}[htbp]%
\vbox{ \hfil \scalebox{1.0}{ {\includegraphics{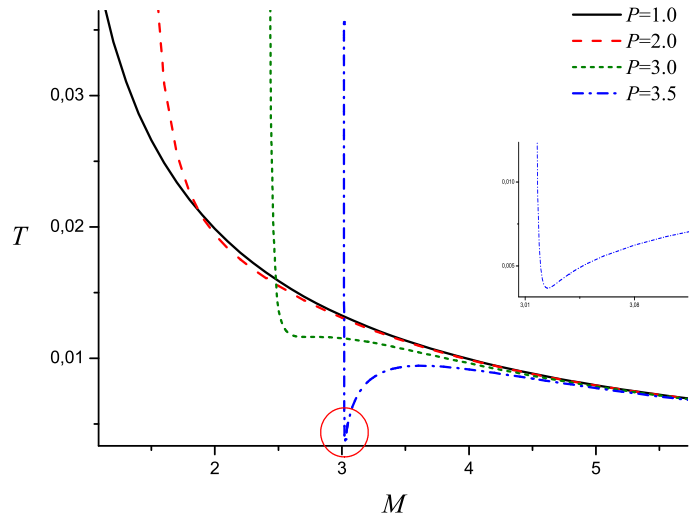}} }\hfil}%
\bigskip%
\caption{%
%------------------------------
$ M$--$T $ relation for the same value of the parameters
as in Fig.(\ref{PhiM}). A magnification of the circled region is also shown. } \label{TMagn}%
\end{figure}%

\end{document}